\documentclass[%
aip,
reprint,%
]{revtex4-1}

\usepackage{graphicx}
\usepackage{dcolumn}
\usepackage{bm}
\usepackage[usenames,dvipsnames]{color}
\usepackage{ulem}
\usepackage{amsthm}
\usepackage{amsmath}
\usepackage{amssymb}
\usepackage{mathrsfs}
\usepackage[disable]{todonotes}
\usepackage{epstopdf}
\usepackage{pdfpages}
\AppendGraphicsExtensions{.tif}

\newcommand{\tn}[1]{\textnormal{#1}}
\newcommand{\eps}{\epsilon}

\newcommand{\epscorr}{\eps_{\tn{cor}}}
\newcommand{\esec}{\eps_{\tn{sec}}}
\newcommand{\Qtol}{\hat{Q}}
\newcommand{\Vtol}{\hat{V}}
\newcommand{\leak}{m_{\rm IR}}
\newcommand{\HminOp}{H_{\min}}

\newcommand{\sket}[1]{{\ensuremath{\lvert#1\rangle}}}
\newcommand{\lket}[1]{{\ensuremath{\left\lvert#1\right\rangle}}}
\newcommand{\ket}[1]{\if@display\lket{#1}\else\sket{#1}\fi}

\newcommand{\sbra}[1]{{\ensuremath{\langle#1\rvert}}}
\newcommand{\lbra}[1]{{\ensuremath{\left\langle#1\right\rvert}}}
\newcommand{\bra}[1]{\if@display\lbra{#1}\else\sbra{#1}\fi}
\newcommand{\A}{\ket{\alpha_0}}
\newcommand{\B}{\ket{\alpha_1}}
\newcommand{\T}{\ket{\alpha_t}}

\begin{document}

\preprint{}

\title{Provably Secure and Practical Quantum Key Distribution over 307 km of Optical Fibre}

\author{Boris Korzh}
 \email{boris.korzh@unige.ch.}
\author{Charles Ci Wen Lim }
 \email{ciwen.lim@unige.ch}
\author{Raphael Houlmann}
\author{Nicolas Gisin}
\affiliation{ Group of Applied Physics, University of Geneva, Geneva, Switzerland}
\author{Ming Jun Li}
\author{Daniel Nolan}
\affiliation{Corning Incorporated, New York, United States}
\author{Bruno Sanguinetti}
\author{Rob Thew}
\author{Hugo Zbinden}
\affiliation{ Group of Applied Physics, University of Geneva, Geneva, Switzerland}


\maketitle
\section{Introduction}
Proposed in 1984, quantum key distribution (QKD) allows two users to exchange provably secure keys via a potentially insecure quantum channel~\cite{bb84}.~Since then, QKD has attracted much attention and significant progress has been made in both theory and practice~\cite{Gisin2002, Scarani2009}.~On the application front, however, the operating distance of practical fibre-based QKD systems is limited to about 150 km~\cite{lo2014}, which is mainly due to the high background noise produced by commonly used semiconductor single-photon detectors~\cite{Hadfield2009, eisaman2011} (SPDs) and the stringent demand on the minimum classical-post-processing (CPP) block size~\cite{Scarani2008, tomamichel2012, lim2014}.~Here, we present, for the first time, a compact and autonomous QKD system that is capable of distributing provably-secure cryptographic key over 307\,km of ultra-low-loss optical fibre (51.9\,dB loss). The system is based on a recently developed standard semiconductor (inGaAs) SPDs~\cite{korzh2014a} with record low background noise and a novel efficient finite-key security analysis for QKD. This demonstrates the feasibility of practical long-distance QKD based on standard fibre-optic telecom components.

In order to achieve long-distance QKD, existing systems usually resort to using superconducting nanowire single-photon detectors (SNSPD) which can achieve a low dark count rate (DCR), but require cryogenic temperatures (\textless3~K). In addition, a weaker security framework is usually taken, i.e., by assuming individual attacks instead of coherent attacks. Crucially, all record-distance demonstrations to date have ignored corrections due to finite-length keys~\cite{Scarani2008, tomamichel2012, lim2014}; in particular, it has been shown that corrections due to finite-length keys are non-negligible for realistic CPP block sizes. This means that previous QKD demonstrations might be overly optimistic in the achievable distance.~Table \ref{tab1} summarises recent fibre based QKD demonstrations. For distances beyond 160\,km, the systems required the use of SNSPDs and thus are incompatible with compact implementations. To the best of our knowledge, only a handful of QKD implementations take finite-length key corrections into account for their security analyses. For example, see the demonstration by Lucamarini {\it et al}~\cite{lucamarini2013}, which distributed provably-secure keys using a fibre of about 90\,km. Although we have restricted our comparison to fibre based prepare-and-measure discrete-variable experiments, we note that significant progress has also been made in continuous variable \cite{jouguet2013} and free-space QKD \cite{nauerth2013, wang2013}.

\begin{table*}
\caption{\label{tab1} Summary of notable QKD demonstrations, outlining the maximum achievable distance, detector type used, and the level of security. Where the finite-key analysis was not carried out the CPP block size, $n_\tn{cpp}$, was left blank and the security parameter, $\epsilon_\text{qkd}$ could not be defined.}

\begin{ruledtabular}
\begin{tabular}{cccccccccc}
&\multicolumn{2}{c}{Quantum channel}&\multicolumn{2}{c}{Detector}&&\multicolumn{3}{c}{Security}\\ \hline
& Length&Attenuation& Type & Temperature & Protocol & Attacks & Finite-key size & $\epsilon_\text{qkd}$ & $r_\text{sec}$ \\ 
& (km) & (dB) & & (K) & & & & & (bps) \\ \hline
This work & 307 & 51.9 & InGaAs~\footnotemark[3] & 153 & COW & Collective & 6.6$\times 10^5$ &$4 \times 10^{-9}$ & 3.18 \\
Wang 2012 \cite{wang2012}& 260 & 52.9 & SNSPD & 1.7 & DPS & Individual & -   & - & 1.85 \\ 
Stucki 2009 \cite{stucki2009}& 250 & 42.9 & SNSPD & 2.5 & COW & Collective &  -  & - & 15 \\
Takesue 2007 \cite{takesue2007}& 200 & 42.1 & SNSPD & 3 & DPS & Individual & - & - & 12.1 \\
Liu 2010 \cite{liu2010}&200& - & SNSPD & 2.4 & BB84 & Collective & - & - & 15 \\
Rosenberg 2009 \cite{rosenberg2009} & 135 & 27.8 & SNSPD & 3 & BB84 & Collective & - & - & 0.2 \\  
Namekata 2011 \cite{namekata2011}& 160 & 33.6 & InGaAs~\footnotemark[4] & 193 & DPS & Individual & - & - & 490 \\
Yuan 2009 \cite{yuan2009} & 100 & 20 & InGaAs~\footnotemark[4] & 243 & BB84 & Collective & - & - & $1.01\times10^4$ \\
Shimizu 2014 \cite{shimizu2014}& 90~\footnotemark[1]  & 30 & SNSPD & 2.5 & DPS & Individual & - & - & 1100\\
Lucamarini 2013 \cite{lucamarini2013} & 80~\footnotemark[2] & 16 & InGaAs~\footnotemark[4] & 243 & BB84 & Collective &$\sim 10^{9}$ & $\sim 10^{-10}$ & $1.20 \times 10^{5}$\\
Walenta 2014 \cite{walenta2014}& 25~\footnotemark[2] & 5.3  & InGaAs~\footnotemark[4] & 293 & COW & Collective & $10^{6}$ & $4 \times 10^{-9}$ & $2.25 \times 10^4$ \\ 
\end{tabular}
\end{ruledtabular}
\footnotetext[1]{Installed fibre link}
\footnotetext[2]{Wavelength multiplexing of classical and quantum channels over a single fibre}
\footnotetext[3]{Free-running operation}
\footnotetext[4]{Gated operation}
\end{table*}
\section{Results}
\begin{figure*}
\includegraphics[width=17cm]{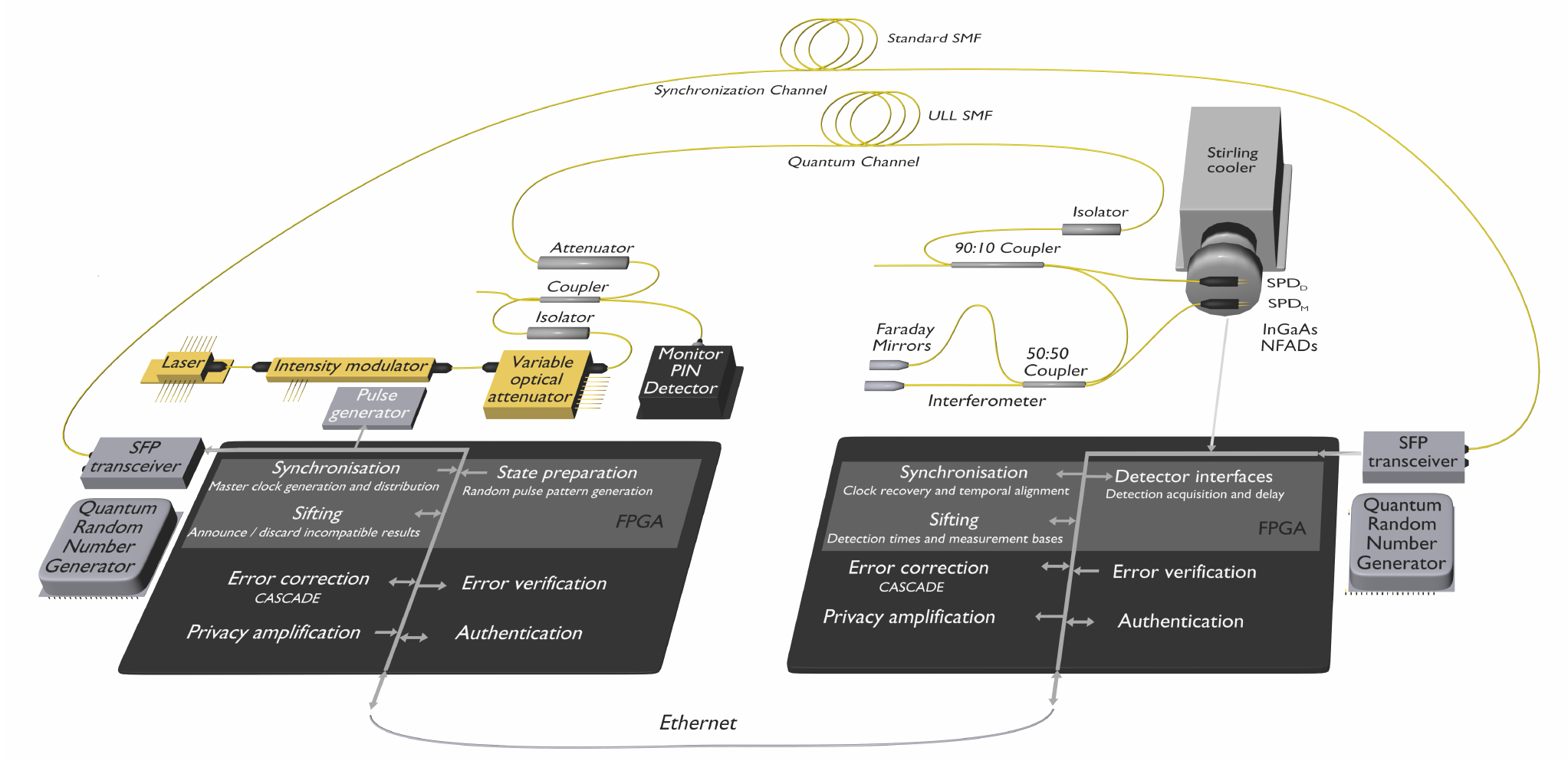} 
\caption{\label{fig:setup} Experimental setup of the COW QKD system. Alice's part (left) consists of a continuous wave DFB laser at 1550\,nm which is modulated using an electro-optic intensity modulator, before passing through a set of attenuators to regulate the photon number per pulse. The state preparation frequency is set to 625\,MHz. Bob's system consists of an asymmetric beam splitter which provides a passive choice of measurement basis, following which the photons either travel directly to the $\text{SPD}_\text{D}$ (Data detector) or pass through an imbalanced interferometer. The phase of the interferometer is maintained such that $\text{SPD}_\text{M}$ is on the destructive port, and the visibility can be calculated by registering the detections due to the interfering (two neighbouring pulses) and non-interfering events (single pulse). The two InGaAs/InP NFAD SPDs are both cooled by a single stirling cooler. The quantum channel consists of ULL fibre whilst a clock signal is distributed from Alice to Bob via the synchronisation channel, both of which are equal in length. All of the processes from the state preparation through to the sifting are carried with the use of a field-programable gate array (FPGA) on each side, whilst the subsequent CPP (error correction, etc) is completed in on-board software.~An ethernet link is used as the service channel and all of the passed messages are authenticated.}
\end{figure*}

Our system is based on the coherent one-way (COW)~\cite{stucki2005} QKD protocol, where the bit string is encoded in the time of arrival of weak coherent laser pulses (WCPs) and the channel disturbance is monitored by measuring the visibility of the interference between neighbouring pulses.  That is, bit~0~and~1~are sent using $\A:=\ket{0}\ket{\alpha}$ and $\B:=\ket{\alpha}\ket{0}$, respectively. On Bob's side, he simply recovers the bit value by measuring the arrival time of the laser pulse, e.g., bit 1 is detected if there is a detection in the later time. To detect attacks on $\A$ and $\B$,~Alice randomly sends an additional test state,~$\T:=\ket{\alpha}\ket{\alpha}$,~to check for phase coherence between any two successive laser pulses. Therefore, phase coherence can be checked in any of these sequences,~$\A\B$, $\A\T$, $\T\B$, $\T$, $\T\T$,~by using an imbalanced interferometer with a pulse delay on Bob's side. The physical implementation is outlined in figure~\ref{fig:setup}. 

The security of our QKD system is based on the universally composable security framework~\cite{RennerThesis2005, Quade2009}.~In particular, we say that our QKD is $\eps_{\tn{qkd}}$-{secure} whenever it is $\esec$-{secret} and $\epscorr$-{correct}. Here, $\esec$-secret means that the output secret key is distinguishable from an ideal secret key with probability at most $\esec$, and $\epscorr$-correct means that probability of Alice and Bob having identical secret keys is at least $1-\epscorr$. Using this security framework and under the assumption of collective attacks~\cite{Branciard2008}, we are able to derive a bound on the maximum value of the extractable secret key length, $\ell$, in terms of a fixed security parameter $\eps_{\tn{qkd}}$ and the observed statistics. \todo{We note that in Ref~\cite{Moroder2012}, it has been shown that there is only a small advantage for the adversary if she uses general attacks instead of collective attacks. }Accordingly, this allows us to select the appropriate family of universal hash functions for privacy amplification, which extracts a secret key of size $\ell$ from a weakly random string of size $n_\tn{cpp}$. Specifically, for some parameter $\beta \in (0,\eps_{\tn{qkd}}/4)$, laser pulse intensity $\mu$, CPP block size, $n_{\tn{cpp}}$, and the number of bits revealed during error correction, $\leak$, we can extract a $\eps_{\tn{qkd}}$-secure key of length (see Supplementary Information (SI) for the full security analysis)
\begin{multline}\label{Main_Eq}
\ell \leq  \max_{\beta}\Bigg\lfloor n_\tn{cpp}\left[1-\Qtol-(1-\Qtol)h\left(\frac{1+\xi(\mu,\Vtol}{2}\right)\right] \\
-7\sqrt{n_\tn{cpp}\log_2 \frac{1}{\beta}}-\leak-\log_2\frac{2}{4\epscorr\beta^2} \Bigg \rfloor, \end{multline}
where $h(x)$ is the binary entropy function and $\xi(a,b):=(2\Vtol-1)\exp(-\mu)-2\sqrt{(1-\exp(-2\mu))\Vtol(1-\Vtol)}$. Here, $\hat{Q}$ and $\hat{V}$ are the measured quantum bit error rate (QBER) in the raw string and the visibility, respectively. 

\begin{figure}
\includegraphics[width=9cm]{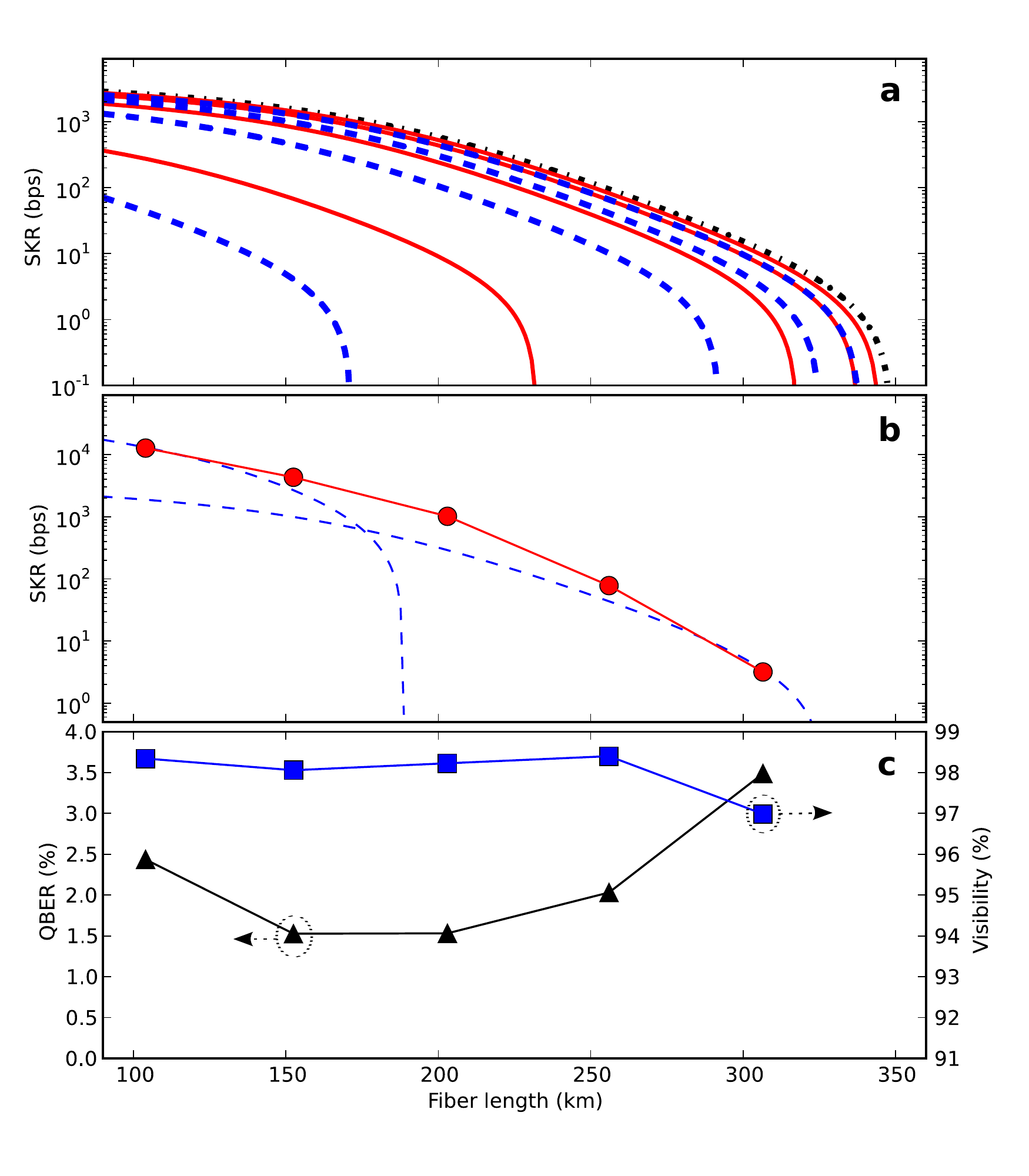}  
\caption{\label{fig:skr}~(\textbf{a}) Numerical optimisation of the SKRs versus distance for different CPP block sizes $n_\tn{cpp}=10^s$ with $s = 4, 5, 6, 7$ (left to right). The results are calculated using the old bound~\cite{walenta2014} (dashed blue lines) and the bound presented in this work (solid red lines). The asymptotic limit is represented by the black dash-dot line. Experimental parameters used for the calculation were taken as for the 307~km experimental measurement and the visibility was set to 98.0\%.~(\textbf{b}) Experimental final secret key rate versus distance. Theoretical plots for different temperature limits are plotted, which show the advantage of increasing the temperature at shorter fibre lengths, which reduces the detector saturation.~(\textbf{c}) QBER (black triangles) and visibility (blue squares) measured for each fibre length.}
\end{figure}

In our system, the QBER is directly measured by counting how many bits are flipped during the error corruption step, meaning that, provided the verification step passes, the uncertainty in this measurement is zero. For the estimation of visibility, we need to consider the problem of random sampling without replacement, i.e., based on the observed visibility $V_{\tn{obs}}$ in the monitoring line, we want to estimate the visibility $V_{\tn{key}}$ of the quantum signals used to generate the secret key. More formally, for some positive deviation term $t$, we like to show that $V_{\tn{key}}\leq V_{\tn{obs}}-t$ is highly unlikely; note that $\hat{V}=V_{\tn{obs}}-t$. In the literature, finite-key security analyses commonly use the Chernoff-Hoeffding tail inequality or Serfling tail inequality to solve this problem. However, these inequalities are only optimal for sufficiently large sample sizes, and thus are not suitable for long-distance QKD where sample sizes are small. To resolve this issue, we derive a new tail inequality that is specifically tailored for random sampling without replacement. In particular, we exploit the fact that the distribution of errors in a random sample is described by the hypergeometric distribution, and then use tight bounds for binomial coefficients to derive an upper bound on $\Pr\left[   V_{\tn{key}}\leq V_{\tn{obs}}-t  \right]$. More precisely, we are able to show that the following relation,
\begin{equation}\label{Main_Eq2}
V_{\tn{key}} \overset{}{\geq} V_{\tn{obs}}-t (n_{\tn{cpp}},n_\tn{vis}, V_{\tn{obs}},\beta),
\end{equation}holds with probability at least $1-\beta$, and where $n_{\tn{vis}}$ is the number of events used to calculate $V_{\tn{obs}}$. The explicit expression of $t (n_{\tn{cpp}},n_\tn{vis}, V_{\tn{obs}},\beta)$ is deferred to the Methods section. In comparison to existing tail inequalities, the new inequality takes into account the measured error rate of the random sample and thus provides a much sharper bound on the tail event, i.e.,$V_{\tn{key}}\leq V_{\tn{obs}}-t  $. To illustrate the tightness of this new bound, figure~\ref{fig:skr}a shows the numerically optimised secret key rates (SKRs) for different CPP block sizes based on the previous bound~\cite{walenta2014} and the new tail inequality. The performance enhancement with small $n_{\tn{cpp}}$ is significant, allowing key distribution to around 310\,km even with $n_{\tn{cpp}}=10^5$, which would not have been possible with the old bound. At 300\,km the SKR reaches 84.5\% of the asymptotic limit with $n_{\tn{cpp}}=10^7$, which is an order of magnitude smaller than previously thought required~\cite{lucamarini2013}.

Another enabling factor of this work has been our recent demonstration that detectors based on InGaAs/InP negative feedback avalanche diodes (NFADs)~\cite{Itzler2011} can operate with extremely low noise in the free-running regime~\cite{korzh2014a}. This marked an improvement of over two orders of magnitude compared to previous InGaAs detectors. Due to a separate avalanche and multiplication~\cite{Itzler2011} structure typically used in such SPDs, where InGaAs is used for the absorption region, and InP for the avalanche region, dark carriers are generated by two different mechanisms. In fact, at temperatures above 200\,K, thermal dark carrier generation in the absorption region is dominant: below this temperature, trap-assisted tunnelling (TAT) in the multiplication region becomes the dominant effect. Generally, TAT has a small temperature dependence, however, the breakdown voltage ($V_{\text{BR}}$) of InP has a linear dependance on temperature. This means that reducing the temperature below 200\,K still leads to a reduction of the DCR, because the operating voltage is reduced, lowering the TAT contribution which is a field-dependent effect~\cite{korzh2014b}. The NFADs used in this report have a ($V_{\text{BR}}$) temperature coefficient such that below 200\,K the DCR drops by approximately a factor of two with every 10\,K. We have observed this trend down to temperatures of less than 150\,K, where a DCR of a few counts per second can be achieved at detection efficiencies of more than 20\%, which is comparable to many SNSPDs~\cite{eisaman2011}.

\begin{figure}
\includegraphics[width=9cm]{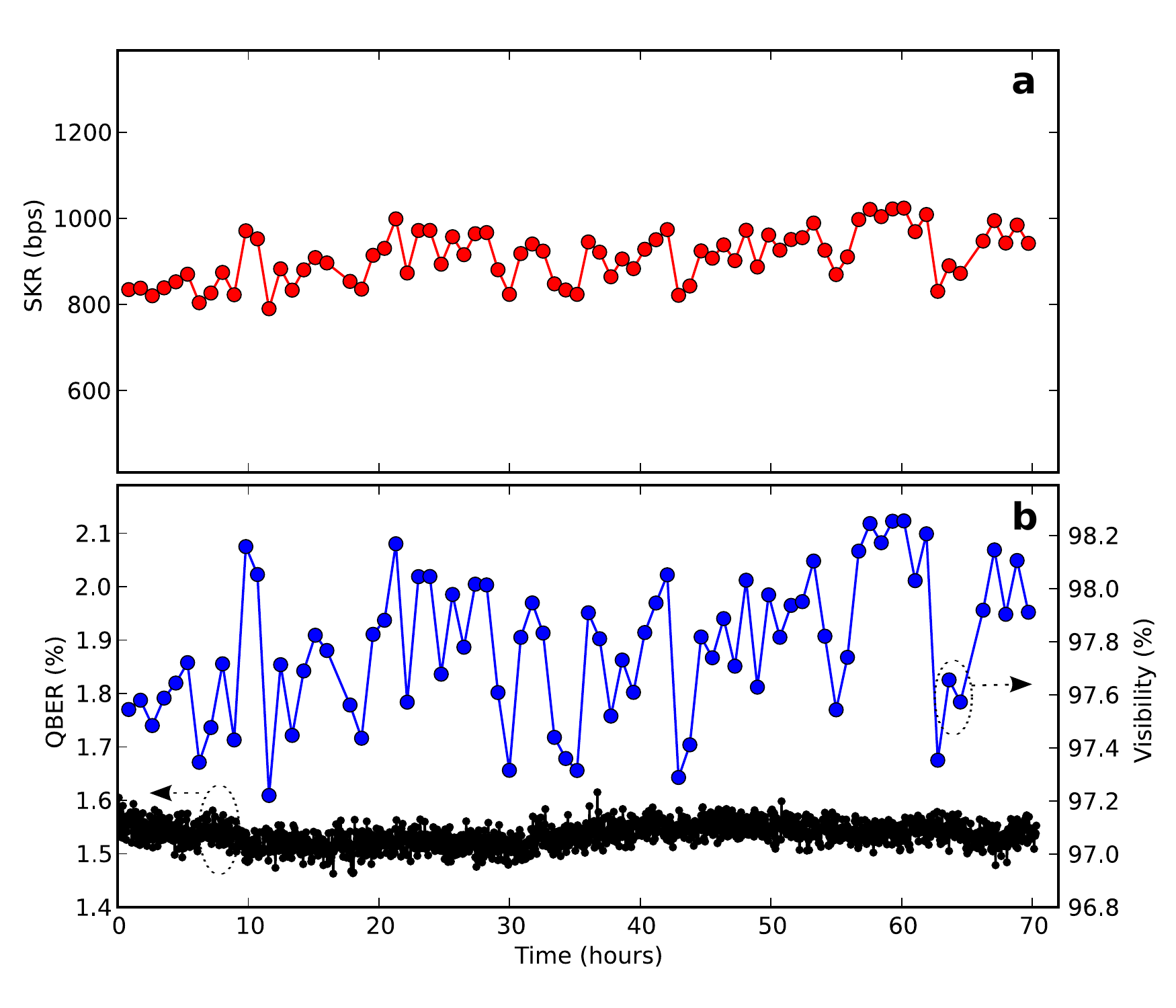}
\caption{\label{fig:stability} QKD system stability over 64 hours with 200\,km of fibre showing~(\textbf{a}) the SKR and (\textbf{b}) the corresponding QBER and visibility as a function of time. The detector temperature was set to 183 K. one of the advantages of the COW protocol is that the bit string is encoded in the time basis and the phase basis is only used for the monitoring of the channel disturbance. Hence, even if the visibility is low, the errors in the raw bit string are not necessarily affected, meaning only the eavesdroppers mutual information is altered but not the $\leak$ in Eq.~\ref{Main_Eq}. This is contrary to the DPS protocol \cite{shimizu2014}, where the bit string is encoded in the phase, such that both of the terms would be affected. As seen in (\textbf{a}), the QBER fluctuations are significantly smaller than those of the visibility, highlighting the importance of this.}
\end{figure}

In practice it is not always beneficial to operate the detector at the lowest temperature since, for a given dead-time, the after-pulse probability increases exponentially with reducing temperatures~\cite{korzh2014b}.~This is not a problem because as the channel length is increased (i.e., increasing channel loss), the detection rates drop exponentially, allowing a longer dead-time to be applied without effecting the detection rates. Therefore, for each distance there exists an optimum detector temperature, which should be set such that the DCR is close to being the dominant source of the QBER. 

The QKD system was tested over fibre lengths of 100-307\,km and the operating temperature was varied between 223-153\,K, whilst the dead time was varied between 8-115~$\mu s$\,(see SI for details of the experimental settings).~The SKRs achieved at each distance are shown in figure~\ref{fig:skr}b and the corresponding error rates are plotted in figure~\ref{fig:skr}c. Theoretical curves are plotted for the minimum and maximum detector temperatures to demonstrate the adaptability of the system at different distances.\,A SKR of 12.7\,kbps was generated at 104\,km, while at 307~km the SKR was 3.18~bps. Since the DCR increases faster with temperature above 200\,K~\cite{korzh2014b}, it was not feasible to increase the temperature to more than 223\,K, where a minimum achievable dead time was around 8\,$\mu$s. For this reason, the NFAD SPDs are optimal for distances $\textgreater$100\,km.\,For shorter distances, it is better to use rapid gating detectors operating as high as room temperature, which do not require dead time \cite{walenta2012}.~At the longest distance, the visibility dropped to $97.0\%$ (from $\textgreater 98\%$) due to the increased difficulty of its stabilisation (see Methods), due to the large integration times required. We adapted the $n_\tn{cpp}$ for each fibre length to maintain reasonable collection times (see SI for details) and could use $n_\tn{cpp}=6.6\times10^5$ even at the longest distance. Figure~\ref{fig:stability} shows the system performance over a continuous period of 70\,hours at a distance of 200\,km, showing that stable operation could be maintained, with automatic tracking of the temporal alignment and the visibility. The average QBER and visibility were $1.55\%$ and $97.7\%$ respectively whilst the SKR was around 900\,bps.   

\section{Conclusion}
To summarise, we have demonstrated a robust and autonomous QKD system, based on practical and compact InGaAs SPDs, which achieves secure key distribution over 307~km.~Moreover, we have sharpened the finite-key security analysis, which improves the performance of the QKD protocol when using small CPP block sizes. This has enabled us to provide a quantifiable security parameter for the complete protocol ($\eps_{\tn{qkd}} = 4\times10^{-9}$), which has not been possible before for QKD systems operating over 150\,km.~This work demonstrates that practical, robust and autonomous QKD is feasible over very long distances even with standard telecom components in a rack mounted architecture.\newline
\section{\label{sec:methods}Additional Information}

\textbf{Experimental details} The secret key post-processing flow is illustrated in figure~\ref{fig:setup}.~The information reconciliation (IR) is carried out using the CASCADE algorithm~\cite{Brassard1994}, which was chosen over an LDPC solution~\cite{walenta2014} due to its increased efficiency of processing small blocks and the fact that high throughput was not required. The IR processing block size was typically 2-3 orders of magnitude smaller than $n_{\tn{cpp}}$, meaning that a frequent measure of the QBER could be obtained, which is useful for active stabilisations of parameters such as the intensity modulator bias voltage. Once the error corrected string reaches the size of the CPP block, error verification is carried out, followed by computation of equation~(\ref{Main_Eq}) to set the compression ratio of the subsequent privacy amplification (PA) step. We operate in a trusted detector scenario, meaning that the dark count contribution to $\hat{V}$ and $\hat{Q}$ in equation~(\ref{Main_Eq}) are subtracted. To facilitate this, only the detector DCR has to be determined accurately, however, it is straightforward to characterise, even at random times. The security level in this work was chosen such that $\beta = 10^{-9}$,~whilst the failure probabilities of the error verification and service channel authentication were $\sim10^{-11}$ and $\sim10^{-15}$.~This gives us an upper bound on the total security parameter for our system of $\eps_{\tn{qkd}} = 4 \times 10^{-9}$,~for all fibre lengths tested in this work.

To achieve long term operational stability, the fibre length mismatch between the synchronisation and quantum channels is tracked by oversampling the input detector signal at 2.5\,GHz and using the subsequent statistics to shift the detection window even before significant errors are induced. The visibility specifically, was stabilised by adjusting the wavelength of the DFB laser through the adjustment of the drive current.

The SPD temperatures were achieved using a stirling cycle cooler, which are significantly more efficient than electrical Peltier coolers and are capable of achieving significantly lower temperatures. Such cooling technology will enable the complete QKD system to be integrated into telecom standard rack mountable package.

\textbf{Sketch of security analysis.~}
Here, we briefly sketch the proof for the bound on the secret key length, i.e., equation~(\ref{Main_Eq}), and the complete security analysis is deferred to the SI. The technical part of our security analysis lies in finding a bound on the smooth min-entropy of ${X}$ given ${E}$, i.e., $\HminOp^{\eps}({X}|{E})$, where $\eps$ is the smoothing parameter, ${X}$ is the random variable describing the raw key and ${E}$ represents the overall knowledge of the adversary. To arrive at equation~(\ref{Main_Eq}), we first use the quantum leftover hash lemma~\cite{RennerThesis2005}, which says that the extractable secret key length $\ell$ is approximately equal to the $\HminOp^{\eps}({X}|{E})$. 

Second, by using certain chain rules for smooth entropies and assuming collective attacks, we are able to put a lower bound on the smooth min-entropy of ${X}$ given ${E}$ in terms of the von Neumann entropy of ${X}$ given ${E}$. This essentially allows us to use the main results from Branciard {\it et al}~\cite{Branciard2008}., where they found a lower bound on the von Neumann entropy of ${X}$ given ${E}$ for the COW protocol in terms of the expected bit error rate and visibility. In order to account for finite-size correction of the visibility statistics, we derive a tight tail inequality that allows us to relate the expected visibility $V_\tn{key}$ to the observed $V_{\tn{obs}}$. More specifically, let $\lambda=1/2-V_{\tn{obs}}/2$, then the deviation term in Eq.~(\ref{Main_Eq2}) (for simplicity we let $t(n_{\tn{cpp}},n_{\tn{vis}},\lambda,\eps)=t$) is given as 
\begin{equation}
t=\sqrt{\frac{8(n_\tn{cpp}+n_\tn{vis})\lambda(1-\lambda)}{n_\tn{vis}n_\tn{cpp}}\log\frac{\sqrt{n_\tn{cpp}+n_\tn{vis}}C}{\sqrt{2\pi n_\tn{cpp}n_\tn{vis}\lambda(1-\lambda)}\eps}}, \end{equation}
where 
\begin{equation}
C=e^{\left(\frac{1}{8(n_\tn{cpp}+n_\tn{vis})}+\frac{1}{12n_\tn{vis}}-\frac{1}{12n_\tn{vis}\lambda+1}-\frac{1}{12n_\tn{vis}(1-\lambda)+1}\right)}.
\end{equation} In other words, $V_{\tn{key}} \geq V_{\tn{obs}}-t=:\hat{V}$ is true except with probability $\eps$. Using this tail inequality together with the above arguments, we thus arrive at an upper bound for $\ell$ which is essentially dependent on the security parameter $\eps_\tn{qkd}$, observed statistics, the CPP block size $n_\tn{cpp}$ and the random sample size $n_\tn{vis}$.

\textbf{Ultra-low-loss fibre.~} Beyond improving the performance of the QKD system, significant gains can be achieved by reducing the loss of the quantum channel. Due to this, developments in optical fibre technology hold at important role for QKD performance. The ultra-low-loss fibres based on silica were achieved by taking careful considerations to the doping levels and the manufacturing process (refer to SI for full discussion). The fibers used in the current work have an average attenuation of 0.160\,dB/km without splices and connectors.~It is expected that it could be possible to create fibers with an attenuation of less than 0.1\,dB/km in the future \cite{tsujikawa2007}, which would mean that the distance of QKD could increase to over 500\,km.  

\section*{Acknowledgements}
We would like to acknowledge Nino Walenta for useful discussions and Olivier Guinnard for technical assistance. We thank ID Quantique for providing the error correction software used in this work as well as Mathilde Soucarros for technical support.~This work was supported by the Swiss NCCR QSIT project.

\newpage


\section*{Supplementary material}

# Supplementary Information: Provably secure and practical quantum key distribution over 307 km of optical fibre

Boris Korzh,[1,*] Charles Ci Wen Lim,[1,†] Raphael Houlmann,[1] Nicolas Gisin,[1] Ming Jun Li,[2] Daniel Nolan,[2] Bruno Sanguinetti,[1] Rob Thew,[1] and Hugo Zbinden[1]

[1] *Group of Applied Physics, University of Geneva, Geneva, Switzerland*
[2] *Corning Incorporated, New York, United States*

## I. EXPERIMENTAL PARAMETERS

In the table below, we list out the key experimental (optimised) parameters for each distance.

TABLE I. Overview of experimental parameters at different distances.

| Quantum channel | | | Detectors | | | | Distillation | | | | | Performance | | |
|---|---|---|---|---|---|---|---|---|---|---|---|---|---|---|
| Length (km) | Attn (dB) | $T$ (K) | $DCR_{dat}$ (cps) | $\tau_{dat}$ ($\mu s$) | $DCR_{mon}$ (cps) | $\tau_{mon}$ ($\mu s$) | $\mu$ | $n_{cpp}$ | $t_{cpp}$ (s) | $f_{ec}$ | $f_{sec}$ | $V_{obs}$ | $Q_{tot}$ | $r_{sec}$ (bps) |
| 104 | 16.9 | 223 | 548 | 8.6 | 384 | 33.9 | 0.06 | $1.97 \cdot 10^7$ | 536.8 | 1.217 | 0.347 | 0.983 | 0.024 | $1.27 \cdot 10^4$ |
| 153 | 25.7 | 203 | 117 | 25.4 | 86.54 | 76.0 | 0.09 | $1.97 \cdot 10^7$ | 1450 | 1.272 | 0.310 | 0.981 | 0.015 | $5.20 \cdot 10^3$ |
| 203 | 34.1 | 183 | 16.2 | 42.3 | 14.4 | 88.9 | 0.10 | $1.05 \cdot 10^7$ | 3108 | 1.271 | 0.301 | 0.982 | 0.015 | $1.02 \cdot 10^3$ |
| 256 | 42.6 | 163 | 3.26 | 84.5 | 2.80 | 105.8 | 0.09 | $1.3 \cdot 10^6$ | 8586 | 1.228 | 0.230 | 0.984 | 0.020 | 78.0 |
| 307 | 51.9 | 153 | 1.33 | 114 | 0.87 | 114 | 0.07 | $6.6 \cdot 10^5$ | 17245 | 1.287 | 0.084 | 0.970 | 0.035 | 3.18 |

## II. DETAILS OF SECURITY ANALYSIS

Here, we provide details of the security analysis mentioned in the methods section. First, we introduce the security criteria which we are using. Second, we describe the security model, i.e, the assumptions used in the security analysis, and the class of attacks which we are considering. Finally, we derive a bound on the extractable secret key length in terms of the experimental statistics.

### A. Security Criteria and Universal Composable Security

A QKD protocol either aborts or outputs a secret key pair $S$ and $\hat{S}$ for Alice and Bob, respectively. Here, we assume that the secret key pair $S$ and $\hat{S}$ have the same length $\ell$. That is, the secret key space (given that the protocol does not abort) is the set of all binary strings of length $\ell$. In the event that the protocol aborts, it outputs $S = \hat{S} = \perp$. Ideally, we like the QKD protocol to meet two criteria, namely the *correctness* criterion and *secrecy* criterion. The correctness criterion is satisfied if the secret keys are identical, i.e., $S = \hat{S}$. To state the secrecy criterion, we first need to have a description of the correlation between the secret key and Eve. In particular, let system $E$ be the information that Eve gathers during the execution of the QKD protocol, then the correlation between Alice and Eve can be described by a classical-quantum state $\rho_{SE} = \sum_s P_S(s)|s\rangle\langle s| \otimes \sigma_{E|S=s}$, where $\{\sigma_{E|S=s}\}$ is the set of conditional quantum states held by

---
[*] boris.korzh@unige.ch.
[†] ciwen.lim@unige.ch

Eve. The secrecy criterion is satisfied if the classical-quantum state $\rho_{SE} = U_S \otimes \rho_E$, where $U_S$ is the uniform mixture of all possible secret key values.

In reality, however, perfect correctness and secrecy cannot be achieved. To allow for some errors, we consider the following correctness criterion and secrecy criterion. Particularly, we say that the protocol is $\varepsilon_{\text{cor}}$-correct if the pair of output secret keys are identical with probability at least $1-\varepsilon_{\text{cor}}$, and the protocol is $\varepsilon_{\text{sec}}$-secret if it outputs a secret key $S$ with $\|\rho_{SE} - U_S \otimes \rho_E\|_1 \leq 2\Delta$ and $(1-p_{\text{abort}})\Delta \leq \varepsilon_{\text{sec}}$, where $p_{\text{abort}}$ is the probability that the protocol aborts. Note that $\|\cdot\|_1$ is the trace norm. In addition, we say that the protocol is $\epsilon$-secure if it is $\varepsilon_{\text{cor}}$-correct and $\varepsilon_{\text{sec}}$-secret with $\varepsilon_{\text{cor}} + \varepsilon_{\text{sec}} \leq \epsilon$.

Importantly, we note that the security definition given above is universal composable [1, 2], in the sense that secret keys generated by a secure QKD protocol can be safely used in other composable cryptographic tasks, e.g., the one-time-pad (OTP) encryption scheme. For example, let $\texttt{KeyGen}_{\text{real}}$ be a QKD protocol that is $\epsilon_1$-secure and $\texttt{Encrypt}_{\text{real}}$ be an encryption scheme that is $\epsilon_2$-secure (i.e., it is $\epsilon_2$-*indistinguishable*[1] from the ideal encryption scheme $\texttt{Encrypt}_{\text{ideal}}$), then universal composability means that the combined crypto-system $\texttt{KeyGen}_{\text{real}} \circ \texttt{Encrypt}_{\text{real}}$ is $(\epsilon_1 + \epsilon_2)$-secure. If $\texttt{Encrypt}_{\text{real}}$ is the OTP encryption scheme, then the security of the cryptosystem is $\epsilon_1$-secure. This means that the ciphertexts generated by the OTP encryption scheme are $\epsilon_1$-indistinguishable from ciphertexts that would have been generated with a perfectly secret key. Similarly, if the authentication scheme used in the QKD protocol is $\epsilon_{\text{auth}}$-indistinguishable from an ideal authentication scheme, then the QKD protocol is simply $\epsilon_{\text{auth}} + \epsilon_1$-secure [3]. This also explains why the security of QKD can be analysed under the assumption of a perfectly secure authentication scheme without any loss of generality.

### B. Coherent One-Way and Security Model

In the recent decade, it has been pointed out by experimentalist working with coherent state QKD systems that the phase coherence between adjacent coherent laser pulses may be useful for detecting photon-number-splitting (PNS) attacks [4]. The first QKD protocol that exploit this property is the differential-phase-shift (DPS) QKD protocol [5], where the bit values are encoded into the phase difference between two consecutive coherent laser pulses, and the receiver on Bob's side is simply an imbalanced interferometer that measures the interference between two consecutive laser pulses. Another similar QKD protocol is the coherent one-way (COW) QKD protocol [6], where instead of encoding the bits into the relative phase of successive coherent laser pulses, the bits are encoded in time. That is, bit 0 and 1 are sent using $|\alpha_0\rangle := |0\rangle|\alpha\rangle$ and $|\alpha_1\rangle := |\alpha\rangle|0\rangle$, respectively. On Bob's side, he simply recovers the bit value by measuring the arrival time of the laser pulse, e.g., bit 1 is detected if there is a detection in the later time. To detect attacks on $|\alpha_0\rangle$ and $|\alpha_1\rangle$, Alice randomly sends an additional test state, $|\alpha_t\rangle := |\alpha\rangle|\alpha\rangle$, to check for phase coherence between any two successive laser pulses. Therefore, phase coherence can be checked in any of these sequences, $|\alpha_0\rangle|\alpha_1\rangle$, $|\alpha_0\rangle|\alpha_t\rangle$, $|\alpha_t\rangle|\alpha_1\rangle$, $|\alpha_t\rangle$, $|\alpha_t\rangle|\alpha_t\rangle$, by using an imbalanced interferometer with a pulse delay on Bob's side

The security of COW protocol has been extensively studied under several attack scenarios [6–9], however, a general security analysis still somewhat remains elusive; note that Ref. [8] gave the most general security analysis to date. This impasse is mainly due to the difficulty in interpreting the phase coherence measurement as a measurement on individual states. Nevertheless, by making reasonable assumptions about Eve's attack strategies, it is possible to derive non-trivial security

---

[1] More precisely, we consider a hypothetical game in which an abstract device called the *distinguisher* has to guess correctly the identity of the protocol (i.e., whether it is the real protocol P or the ideal protocol P*) when given access to the inputs/outputs of the protocol. We say P and P* are $\epsilon$-indistinguishable if, for any distinguisher, the probability of guessing correctly is at most $1/2 + \epsilon/2$.

bounds for the COW protocol. For instance, if Eve is restricted to collective attacks (i.e., Eve interacts identically and independent with each individual state) and that she either forwards an empty state or a single photon to Bob, upper bounds on the asymptotic secret key rate can be derived [7]. In particular, we refer to Ref. [7]'s collective attack model:

$$
\begin{aligned}
|\sqrt{\mu}\rangle|0\rangle &\to |00\rangle|v_{\mu 0}\rangle_E + \sqrt{(1-Q)\mu t}|10\rangle|p_{\mu 0}^{10}\rangle_E + \sqrt{Q\mu t}|01\rangle|p_{\mu 0}^{01}\rangle_E \\
|0\rangle|\sqrt{\mu}\rangle &\to |00\rangle|v_{0\mu}\rangle_E + \sqrt{(1-Q)\mu t}|01\rangle|p_{0\mu}^{01}\rangle_E + \sqrt{Q\mu t}|10\rangle|p_{0\mu}^{10}\rangle_E \\
|\sqrt{\mu}\rangle|\sqrt{\mu}\rangle &\to |00\rangle|v_{\mu\mu}\rangle_E + \sqrt{(1-Q)\mu t}(|01\rangle|p_{\mu\mu}^{01}\rangle_E + |10\rangle|p_{\mu\mu}^{10}\rangle_E),
\end{aligned}
\tag{1}
$$

where $|v_{jk}\rangle \; \forall \; j,k \in \{0,\mu\}$ are the states Eve adds to the vacuum component of each individual two-mode coherent state, and $|p_{jk}^{10}\rangle$ and $|p_{jk}^{01}\rangle \; \forall \; j,k \in \{0,\mu\}$ are the states Eve adds to the single-photon component of each individual two-mode coherent state. This attack model is, in fact, based on the observation that all practical intensity modulators have finite extinction ratio. That is, instead of producing perfectly modulated intensities $\mu$ and 0, practical intensity modulators produce $(1-Q)\mu$ and $Q\mu$. Therefore, we expect $Q$ fraction of the detected events to be erroneous. Finally, we remark that the assumption of collective attacks is not overly optimistic, since it has been shown in Ref. [8] that Eve gains only very little advantage when the most general attack is considered.

### C. Bounds on extractable secret key length

To start with, let us introduce a fundamental result in quantum cryptography, in particular, the *Quantum Leftover Hash Lemma* (QLHL) [1, Corollary 5.6.1], which plays a central role in privacy amplification. The QLHL says that a $\Delta$-secret key of length $\ell$ can be extracted from $X$ (recall that X is the bit string used for key distillation) if

$$\Delta = 2\epsilon + \frac{1}{2}\sqrt{2^{\ell - H_{\min}^\epsilon(X|E^+)}}, \tag{2}$$

where $H_{\min}^\epsilon(X|E^+)$ is *smooth min-entropy* of $X$ given $E^+$, which quantifies the amount of uncertainty in $X$ when given access to quantum side information $E^+$. Here, the quantum side information $E^+$ is the overall information Eve gathers up to the error verification step. The goal here is to show that the proposed secret key length (i.e., Eq. (1) in main text)

$$\ell \leq \max_{\beta \in (0, \varepsilon_{\mathrm{qkd}}/4]} \left\lfloor n_{\mathrm{cpp}} \left[1 - \hat{Q} - (1-\hat{Q})h\left(\frac{1 + \xi(\mu, \hat{V})}{2}\right)\right] - 7\sqrt{n_{\mathrm{cpp}} \log_2 \frac{1}{\beta}} - m_{\mathrm{IR}} - \log_2 \frac{2}{4\varepsilon_{\mathrm{cor}} \beta^2} \right\rfloor, \tag{3}$$

implies a $\Delta \leq \varepsilon_{\mathrm{qkd}}$ secret key. This is equivalent to showing that the smooth min-entropy term is lower bounded by

$$H_{\min}^\epsilon(X|E^+) \geq n_{\mathrm{cpp}} \left[1 - \hat{Q} - (1-\hat{Q})h\left(\frac{1}{2} + \frac{\xi(\mu, \hat{V})}{2}\right)\right] - 7\sqrt{n_{\mathrm{cpp}} \log_2 \frac{1}{\varepsilon}} - m_{\mathrm{IR}} - \log_2 \frac{2}{\varepsilon_{\mathrm{cor}}}, \tag{4}$$

where $\xi(\mu, \hat{V})$ is some function to be defined and explained later.

To do that, first, we use a chain rule for smooth min-entropies [1, Theorem 3.2.12] to get

$$H_{\min}^\epsilon(X|E^+) \geq H_{\min}^\epsilon(X|E) - m_{\mathrm{IR}} - \log_2 \frac{2}{\varepsilon_{\mathrm{cor}}}.$$

Roughly speaking, this says that the side information $X^+$ is simply a concatenation of information $X$ and $m_{\text{IR}} + \log_2(2/\varepsilon_{\text{cor}})$, where the latter is the number of bits revealed over the public channel during error correction and error verification. Then, under the assumption that Eve interacts independently and identically with each individual two-mode coherent state, the smooth min-entropy of $X$ given $E$ can be lower bounded by using [1, Theorem 3.3.4] and [7, Eqs. (15) and (B.8)]. In particular, we get

$$H_{\min}^{\epsilon}(X|E) \geq n_{\text{cpp}} \left[1 - \hat{Q} - (1-\hat{Q})h\left(\frac{1}{2} + \frac{\xi(\mu, \hat{V})}{2}\right)\right] - 7\sqrt{n_{\text{cpp}} \log_2 \frac{1}{\epsilon}}, \tag{5}$$

where for $\exp(-\mu) > 2(\hat{V}(1-\hat{V}))^{1/2}$ and $\hat{V} > 1/2$, we have

$$\xi(\mu, \hat{V}) := (2\hat{V} - 1)\exp(-\mu) - 2\sqrt{\hat{V}(1-\hat{V})(1-\exp(-2\mu))}.$$

Recall from the main text that $\hat{V}$ is the estimate for the visibility measure, which is defined as $\hat{V} := V_{\text{obs}} - t(n_{\text{cpp}}, n_{\text{vis}}, V_{\text{obs}}, \epsilon')$ where $V_{\text{obs}}$ is the observed visibility, $n_{\text{vis}}$ is the number of bits used to calculate $V_{\text{obs}}$ and $\epsilon'$ is the error in the estimation. This estimate lies at the heart of finite-key security analysis for QKD, where statistics of a small random sample drawn from the measurement data is used to estimate the statistics of the remaining bit string. In the language of statistical analysis, this is essentially the same as random sampling without replacement; however, in our case, we do not have access to the population parameters. For most finite-key security analyses, parameter estimation is usually achieved by using tail inequalities belonging to a family of Hoeffding-Chernoff inequalities [10, 11] (see also [12, Theorem 11.2.1]). However, these estimation techniques are often sub-optimal due to the fact that they use only limited information (i.e., the random sample size and population size) to construct the confidence interval. By sub-optimal, we mean that the empirical average has a poor convergent rate (towards the asymptotic average) especially when the sampling size is small. Indeed, put in the context of finite-key security analysis, a small classical post-processing block size would imply an overly pessimistic estimate of the quantum channel error rate. As a result, no key extraction is possible if the underlying block size is small; consequently, this makes QKD very inefficient when the channel loss is high.

To overcome this issue, we consider a direct estimation approach where we exploit the fact that the error sampling distribution (assuming random sampling without replacement) follows the hypergeometric distribution. Then, by making use of a sharp inequality for binomial coefficient, we are able to derive a tight bound on the probability of event $V_{\text{key}} \leq V_{\text{obs}} - t$, where $t$ is some positive parameter to be determined below.

**New tail inequality.** *Let $\mathcal{Z} := \{z_1, z_2, z_3, \cdots, z_{n+k}\}$ be a list of bits, where the number of ones is unknown. Let $\mathcal{Z}_{\text{pe}}$ be a random sample (taken without replacement) of size $k$ of $\mathcal{Z}$, and let $\lfloor k\lambda \rfloor$ be the number of ones observed, where $0 < \lambda \leq 1/2$ is the observed error rate. In addition, let the remaining bits be $\mathcal{Z}_{\text{key}} = \mathcal{Z} \setminus \mathcal{Z}_{\text{pe}}$ with $\lambda_{\text{key}} = \sum_{z_i \in \mathcal{Z}_{\text{key}}} z_i/n$. Then, for some positive parameter $\epsilon' > 0$, we have*

$$\Pr\left[\lambda_{\text{key}} \geq \lambda + g(n, k, \lambda, \epsilon')\right] < \epsilon', \tag{6}$$

*where*

$$g(n, k, \lambda, \epsilon') := \sqrt{\frac{2(n+k)\lambda(1-\lambda)}{kn}} \log \frac{\sqrt{n+k}C(n,k,\lambda)}{\sqrt{2\pi nk\lambda(1-\lambda)}\epsilon'}, \tag{7}$$

$$C(n, k, \lambda) := \exp\left(\frac{1}{8(n+k)} + \frac{1}{12k} - \frac{1}{12k\lambda + 1} - \frac{1}{12k(1-\lambda) + 1}\right). \tag{8}$$

*Proof.* Let random variables, $W_{\text{pe}}$ and $W_{\text{key}}$, be the number of ones in $\mathcal{Z}_{\text{pe}}$ and $\mathcal{Z}_{\text{key}}$, respectively. Then, the number of ones in the list is given by random variable $W = W_{\text{pe}} + W_{\text{key}}$. The goal here is to show that the joint probability of $W_{\text{key}} \geq (n/k)W_{\text{pe}} + nt$ and $W_{\text{pe}} = \lfloor k\lambda \rfloor$, for some small $t > 0$, is extremely unlikely for sufficiently large $n + k$. To do that, we first note that

$$\begin{aligned}
\Pr\left[W_{\text{key}}/n \geq \lambda + t\right] &= \Pr\left[W_{\text{pe}} = k\lambda, W_{\text{key}} \geq n\lambda + nt\right] \\
&= \Pr\left[W_{\text{pe}} = k\lambda, W \geq (n+k)\lambda + nt\right] \\
&= \sum_{w=(n+k)\lambda+nt} \Pr\left[W_{\text{pe}} = k\lambda, W = w\right] \\
&= \sum_{w=(n+k)\lambda+nt} \Pr\left[W_{\text{pe}} = k\lambda | W = w\right] \Pr[W = w] \\
&= \sum_{w=(n+k)\lambda+nt} \frac{\binom{k}{k\lambda}\binom{n}{w-k\lambda}}{\binom{n+k}{w}} \Pr[W = w],
\end{aligned}$$

where the last inequality uses the fact that $\Pr\left[W_{\text{pe}} = k\lambda | W = w\right]$ is given by a hypergeometric distribution: i.e., probability of getting $k\lambda$ ones in $k$ bits drawn randomly without replacement from $n + k$ bits with $w$ ones. Then, by noting that $\Pr\left[W_{\text{pe}} = k\lambda | W = w\right]$ is strictly decreasing for $w > (n+k)\lambda + nt$, we further get

$$\Pr\left[W_{\text{key}}/n \geq \lambda + t\right] \leq \frac{\binom{k}{k\lambda}\binom{n}{n\lambda+nt}}{\binom{n+k}{(n+k)\lambda+nt}} = \frac{\binom{k}{k\lambda}\binom{n}{n\lambda_{\text{key}}}}{\binom{n+k}{(n+k)\lambda_{\text{all}}}}, \quad (9)$$

where $\lambda_{\text{key}} = \lambda + t$ and $\lambda_{\text{all}} = k\lambda/(n+k) + n\lambda_{\text{key}}/(n+t)$. Next, we make use of a sharp double inequality for binomial coefficient [13]:

$$e^{-\frac{1}{8n}} G(\alpha, n) < \binom{n}{\alpha n} < e^{\left(\frac{1}{12n} - \frac{1}{12n\alpha+1} - \frac{1}{12n(1-\alpha)+1}\right)} G(\alpha, n),$$

where for $\alpha \in (0, 1/2]$ and for $n \in \mathbb{N}^+$,

$$G(\alpha, n) := \frac{\alpha^{-\alpha n}(1-\alpha)^{-(1-\alpha)n}}{\sqrt{2\pi n \alpha(1-\alpha)}}.$$

Applying this inequality to equation (9), we get

$$\frac{\binom{k}{k\lambda}\binom{n}{n\lambda_{\text{key}}}}{\binom{n+k}{(n+k)\lambda_{\text{all}}}} < \frac{\exp\left(\log(2)[kh(\lambda) - (n+k)h(\lambda_{\text{all}}) + nh(\lambda_{\text{key}})]\right) C(n,k,\lambda)}{\sqrt{2\pi nk\lambda(1-\lambda)/(n+k)}}, \quad (10)$$

where

$$C(n,k,\lambda) := \exp\left(\frac{1}{8(n+k)} + \frac{1}{12k} - \frac{1}{12k\lambda+1} - \frac{1}{12k(1-\lambda)+1}\right).$$

Note that we used $\sqrt{\lambda_{\text{all}}(1-\lambda_{\text{all}})} \leq \sqrt{\lambda_{\text{key}}(1-\lambda_{\text{key}})}$ (recall that by construction, we have $\lambda_{\text{key}} \geq \lambda_{\text{all}}$), and $\exp(\log(2)h(\alpha)) = (\alpha)^{-\alpha}(1-\alpha)^{-(1-\alpha)}$. Then, by using an inequality for the binary entropy function, i.e., $h(x) \leq h(x_0) + h'(x_0)(x-x_0) + h''(x_0)(x-x_0)^2/2 + h'''(x_0)(x-x_0)^3/6$, we further get

$$\begin{aligned}
nh(\lambda_{\text{key}}) - (n+k)h(\lambda_{\text{all}}) + (k)h(\lambda) &\leq \frac{h''(\lambda_{\text{all}})}{2} \frac{t^2 nk}{n+k} \\
&= -\frac{t^2 nk}{2\log(2)(n+k)\lambda_{\text{all}}(1-\lambda_{\text{all}})} \leq -\frac{t^2 nk}{2\log(2)(n+k)\lambda(1-\lambda)}, \quad (11)
\end{aligned}$$

where the last inequality is obtained by noting that $[x(1-x)]^{-1}$ is a decreasing function for $x \in (0, 1/2]$ and that $\lambda_{\text{all}} \geq \lambda$. Inserting equation (11) into equation (10) gives

$$\frac{\binom{k}{k\lambda}\binom{n}{n\lambda_{\text{key}}}}{\binom{n+k}{(n+k)\lambda_{\text{all}}}} < \frac{\exp\left(-\frac{t^2 nk}{2(n+k)\lambda(1-\lambda)}\right) C(n,k,\lambda)}{\sqrt{2\pi nk\lambda(1-\lambda)/(n+k)}} =: \epsilon',$$

The deviation parameter, $g$, is thus

$$g(n,k,\lambda,\epsilon) := \sqrt{\frac{2(n+k)\lambda(1-\lambda)}{nk}} \log \frac{\sqrt{n+k}C(n,k,\lambda_1)}{\sqrt{2\pi nk\lambda(1-\lambda)}\epsilon'}.$$

□

In order to apply the tail inequality to visibility measure, we make use of the relationship $V = 1 - 2Q$. More specifically, after some simple arrangements, we get $\Pr[V_{\text{key}} \leq V_{\text{obs}} - 2g(n,k,\lambda,\epsilon')] < \epsilon'$. Therefore, the deviation term introduced in the main text is simply $t(n,k,\lambda,\epsilon') = 2g(n,k,\lambda,\epsilon')$. In other words, we have $V_{\text{key}} \geq V_{\text{obs}} - t(n_{\text{cpp}}, n_{\text{vis}}, V_{\text{obs}}, \epsilon') =: \hat{V}$ with probability at least $1 - \epsilon'$. Note that $\epsilon'$ can be seen as the significant level associated with standard confidence interval estimation.

The secrecy of our protocol is obtained by inserting equation (4) into equation (2) and using the proposed secret key length (i.e., equation (3)), which gives $\Delta \leq 2\epsilon + \beta$. Finally, by composing the error probability due to parameter estimation, we thus have a total secrecy of $\Delta \leq 2\epsilon + \beta + \epsilon'$, which may be simplified by choosing $\epsilon = \epsilon' = \beta = \epsilon_{\text{qkd}}/4$; accordingly, we have $\Delta \leq \varepsilon_{\text{qkd}}$.

### III. ULTRA-LOW-LOSS OPTICAL FIBRE

The total attenuation of an optical fibre is the sum of the intrinsic loss factors such as Rayleigh scattering $\alpha_{\text{RS}}$, infrared absorption $\alpha_{\text{IR}}$ and ultraviolet absorption $\alpha_{\text{UV}}$, as well as the extrinsic loss factors such as absorption due to transition metals $\alpha_{\text{TM}}$, absorption due OH ions $\alpha_{\text{OH}}$, scattering due to waveguide imperfections $\alpha_{\text{IM}}$ and loss due to fiber bending effects $\alpha_{\text{BL}}$,

$$\alpha = \alpha_{\text{RS}} + \alpha_{\text{IR}} + \alpha_{\text{UV}} + \alpha_{\text{TM}} + \alpha_{\text{OH}} + \alpha_{\text{IM}} + \alpha_{\text{BL}}. \tag{12}$$

To achieve ultra-low attenuation, it is necessary to minimise the contributions from both extrinsic and intrinsic loss factors. The contaminants due to transition metals can be eliminated practically in the fibre preform manufacturing processes by using chemical vapour deposition techniques with high pure chemical row materials. The OH concentration can be reduced to the minimum level through chlorine drying. Waveguide imperfection loss is caused by the geometry fluctuation at the core and cladding boundary. The boundary fluctuation is mainly due to the residual stress which is induced during the manufacturing process. The residual stress depends on the magnitude of the viscosity difference between the core and cladding and the fibre drawing tension. The stress can be reduced by matching the viscosity of the core and cladding [14]. For the intrinsic factors, the most important one is the Rayleigh scattering loss. The Rayleigh scattering loss can be expressed by the sum of two contributions from density and concentration fluctuations [15],

$$\alpha_{\text{RS}} = \alpha_\rho + \alpha_C. \tag{13}$$

The density fluctuation contribution to the scattering coefficient depends on the fictive temperature, $T_{\text{f}}$, which is determined as the temperature where the glass structure is the same as that of the supercooled liquid,

$$\alpha_\rho = \frac{8\pi^3}{3\lambda^4} n^8 p^2 \beta_{\text{T}} k_{\text{B}} T_{\text{f}}. \tag{14}$$

where $\lambda$ is the wavelength of incident light, $n$ the refractive index, $p$ the photo-elastic coefficient, $k_\text{B}$ the Boltzmann constant, and $\beta_\text{T}$ the isothermal compressibility. The concentration fluctuation contribution is proportional to,

$$\alpha_C \sim \frac{\partial n}{\partial C} \left\langle \Delta C^2 \right\rangle T_\text{f}. \tag{15}$$

Because the Rayleigh scattering is mainly caused by frozen-in density fluctuation, to suppress it as much a possible it is necessary to reduce $T_\text{f}$ to increase structural relaxation. To minimise the concentration fluctuation, it is advantageous to reduce the $GeO_2$ dopant level in the core because the Rayleigh scattering loss is proportional the $GeO_2$ concentration. For this reason, it is better to use pure silica material in the core. However, the pure silica material has a high fictive temperature, which increases the density fluctuation contribution. To reduce the density fluctuation, the core composition needs to be carefully engineered with a small amount of dopant to reduce the fictive temperature while keeping the concentration fluctuation negligible [16, 17]. The fictive temperature is also affected by the cooling rate during the fibre draw. It is advantageous to use a slow cooling rate to promote the structural relaxation. With these considerations, it has been predicted that the intrinsic loss of silica based optical fibers could be below 0.1 dB/km [18].



\end{document}